\documentclass[final,5p]{elsarticle}
\cortext[cor1]{Corresponding author: \texttt{alexander.gregg@newcastle.edu.au}}
\journal{ }
\biboptions{sort&compress}

\usepackage{amsmath,amssymb,amsfonts,amsthm,enumerate,multirow}
\usepackage{placeins}
\usepackage{dblfloatfix}
\usepackage[percent]{overpic}
\usepackage{enumitem}
\usepackage[utf8]{inputenc}

\newcommand{\epsave}{\langle \epsilon \rangle}
\newcommand{\bs}[1]{\boldsymbol{#1}}

\newcommand{\kap}{\boldsymbol{\hat{\kappa}}}
\newcommand{\tL}{\boldsymbol{\mathcal{L}}}

\newcommand{\pd}[2]{\frac{\text{d}#1}{\text{d}#2}}

\newcommand{\Transp}{\mathsf{T}}
\newcommand{\norm}[1]{\left\lVert#1\right\rVert}
\newcommand{\argmax}[1]{\underset{#1}{\operatorname{arg}\,\operatorname{max}}\;}

\begin{document}

\begin{frontmatter}
\title{Neutron Diffraction Strain Tomography: Demonstration and Proof-of-Concept}

\author[NewcE]{A.W.T. Gregg\corref{cor1}}
\author[NewcE]{J.N. Hendriks}
\author[NewcE]{C.M. Wensrich}
\author[ANSTO]{V.Luzin}
\author[NewcE]{A.Wills}
\address[NewcE]{School of Engineering, The University of Newcastle, Callaghan NSW 2308, Australia}
\address[ANSTO]{Australian Centre for Neutron Scattering, Australian Nuclear Science and Technology Organisation, Kirawee NSW 2232, Australia}

\begin{abstract}
Recently, a number of reconstruction algorithms have been presented for residual strain tomography from Bragg-edge neutron transmission measurements. In this paper, we examine whether strain tomography can also be achieved from diffraction measurements. We outline the proposed method and develop a suitable reconstruction algorithm. This technique is demonstrated in simulation and a proof-of-concept experiment is carried out where the strain field in an axisymmetric sample is reconstructed and validated against conventional diffraction strain scans.
\end{abstract}
\end{frontmatter}

\section{Introduction}
For many years, neutron diffraction instruments have been able to provide high-precision measurements of strain (down to $\sigma = 1\times10^{-5}$ uncertainty). by observing changes in the atomic lattice spacing $d$ within a polycrystalline samples \citep{noyan87}. These measurements can be performed with resolution down to $0.5$mm$^3$  based on practical limitations in gauge volume size. 

\begin{figure}[!h]
\begin{center}  \includegraphics[width=\linewidth]{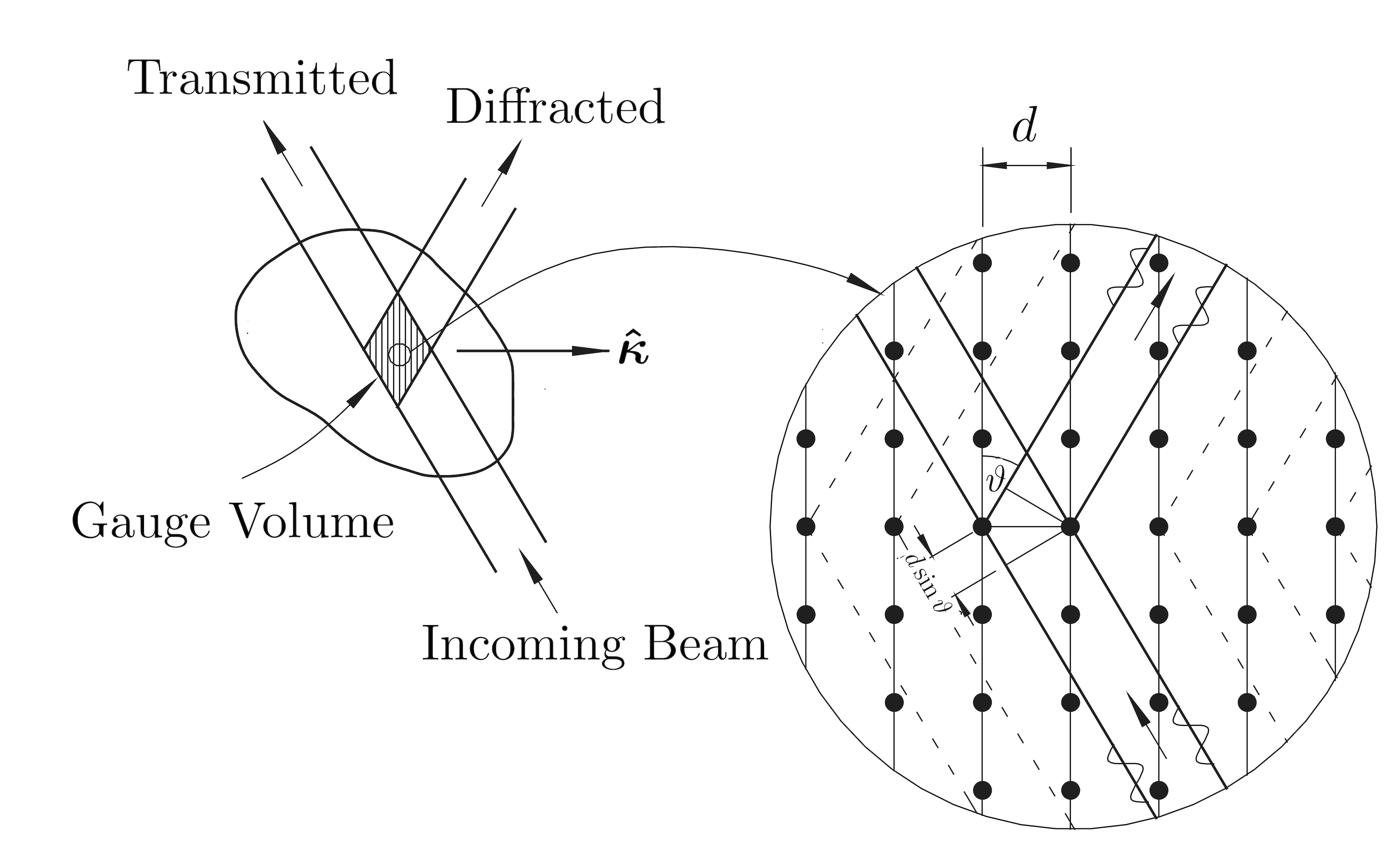}
\vspace{-0.8cm}
\caption{A neutron beam interacts with the lattice of a polycrystalline material \citep{wensrich14}. Some neutrons are diffracted constructively according to Bragg's Law.}
\vspace{-0.5cm}
\label{fig:1} \end{center}
\end{figure}

These measurements rely on Bragg's law, which provides the condition for constructive interference of neutron radiation in a lattice:
\begin{equation*}
\lambda = 2d \sin{\vartheta},
\end{equation*}
where $\lambda$ is the neutron wavelength, $d$ the average spacing of all planes within the gauge volume aligned with the direction $\kap$ bisecting the incident and diffracted beams, and $\vartheta$ the half-angle of diffraction, as shown in Figure \ref{fig:1}.

In conventional neutron strain scanning, $\lambda$ is known and $\vartheta$ measured or vice-versa and the average strain in the $\kap$ direction is then calculated by:
\begin{equation*}
\epsave = \frac{d-d_0}{d_0},
\end{equation*}
where $d_0$ is the lattice spacing in a reference (unstrained) sample.

By taking a number of measurements from different sample orientations, it is possible to resolve the entire triaxial strain tensor at one or a number of points within a body. This `pointwise' approach has obvious application where the area of interest is known; e.g. probing the stress at a fixed location as a function of some external stimulus, or examining the effects of a crack/weld/other feature.

Where full-field strain maps are required (e.g. when evaluating the residual strains locked-in by novel manufacturing techniques), it is commonplace to interpolate between sets of discrete, distributed, point-wise measurements. Such interpolations present two significant issues:

\begin{enumerate}
    \item The quality of the interpolation is dependent on the resolution of the pointwise measurements. Achieving sufficient resolution (particularly in 3 dimensions) is not always practical as beamtime is a limited resource.
    \item Interpolations are not guaranteed to represent the underlying field with a high level of fidelity --- i.e. they may not satisfy physical constraints such as equilibrium or compatibility.
\end{enumerate}

In recent years, a number of algorithms for full-field strain tomography from neutron \emph{transmission} measurements have been developed and have shown promise both in simulation and on experimental data \citep{,abbey09,abbey12,kirkwood15,wensrich16a,wensrich16b,hendriks2017,gregg2017axi,gregg2018resid,hendriks2018traction,jidling2018probabilistic}.

These algorithms rely on Bragg-edge imaging, whereby the the transmitted (as opposed to diffracted) neutrons are counted \citep{santisteban02b}. Presently, only a handful of facilities are suitable for performing strain tomography using these techniques. 

We propose that strain imaging is also possible using diffraction geometry and instrumentation, and that the recently developed reconstruction algorithms can be extended to this problem.

If successful, this will significantly extend the utility of these algorithms in two major ways. First, this may allow strain tomography at the multitude of existing diffraction strain scanners around the world, and secondly, this may open the door for a combined reconstruction algorithm which utilises transmission and diffraction measurements in tandem. This tantalising prospect has the potential to reduce beamtime requirements and improve the quality of strain reconstructions.

In this paper, we present a method and algorithm for strain tomography using diffraction geometry. As a proof-of-concept, we restrict this demonstration to axisymmetric systems. We first demonstrate the approach using simulated measurements, and then reconstruct the strain within a crushed, axisymmetric disc from experimental data. We conclude by briefly discussing potential improvements to the algorithm and the considerations in extending this method to arbitrary 2D systems.

\section{Method}
\subsection{Experimental Setup}

\begin{figure}[!h]
\begin{center}  \includegraphics[width=0.8\linewidth]{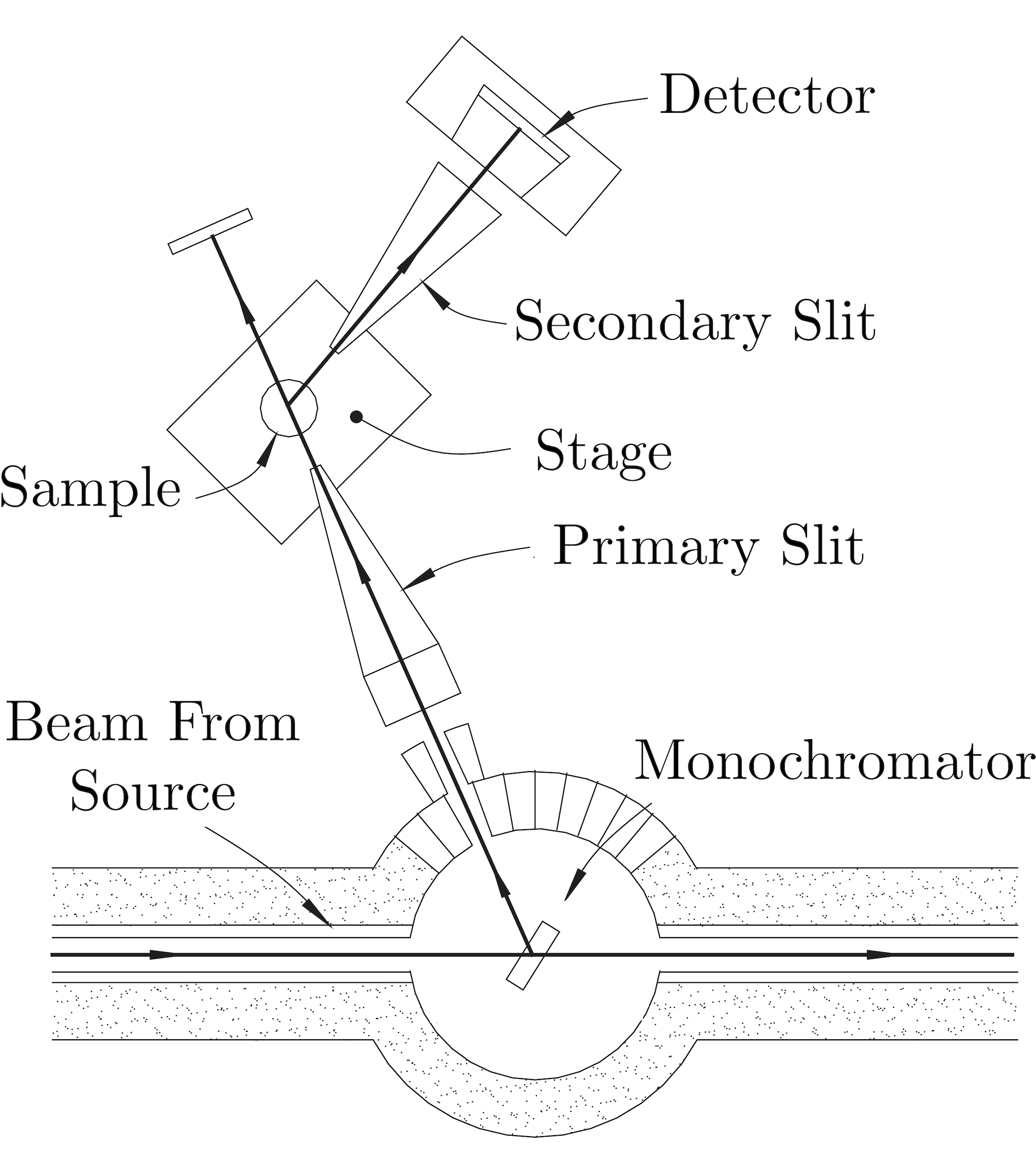}
\vspace{-0.3cm}
\caption{Schematic of a typical constant-wavelength neutron strain scanner. \citep{wensrich14}}
\label{fig:2} \end{center}
\vspace{-0.5cm}
\end{figure}

\begin{figure}[!h]
\begin{center}  \includegraphics[width=0.75\linewidth]{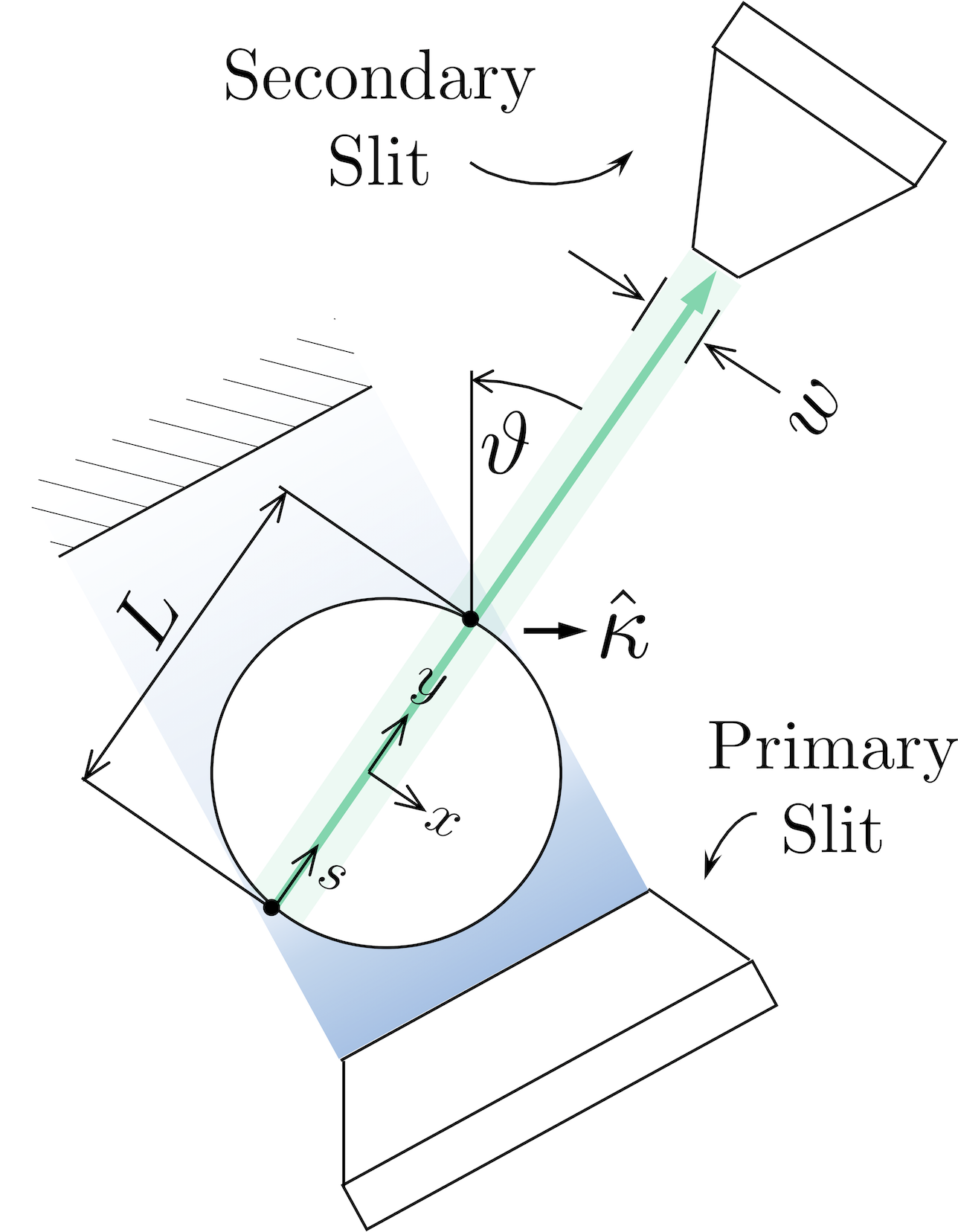}
\vspace{-0.3cm}
\caption{Coordinate System and measurement geometry for diffraction tomography.}
\vspace{-0.8cm}
\label{fig:3} \end{center}
\end{figure}

We propose that conventional diffraction strain scanners can be utilised to obtain `ray-like' measurements and perform strain imaging analogous to a transmission setup. The forward mapping has been examined in \cite{luzin19}. This technique may be feasible for both constant-wavelength (e.g. see Figure \ref{fig:2}) and time-of-flight strain scanners. While the finer details will differ, we provide a general overview for the former:

\begin{enumerate}
    \item A polychromatic beam of neutrons is generated by a reactor source and is directed toward the instrument via a shielded beam line.
    \item This beam is simultaneously redirected, focussed, and reduced to a single wavelength using a curved monochromator.
    \item The monochromatic beam floods the sample (i.e. a fully-open primary slit) and neutrons are scattered outward in various directions by the many, differently oriented planes according to Bragg's law.
    \item A cadmium secondary slit defines an effective gauge volume in the form of a long, thin region that can be approximated as a ray\footnote{Obviously, the secondary slit width $w$ has an effect on the validity of the ray assumption. This is discussed in Section \ref{sec:ext}.} --- see Figure \ref{fig:3}.
    \item Analogous to conventional strain scanning, the measured lattice spacing $d$ then provides the average normal strain in the $\kap$ direction along this ray.
\end{enumerate}

By sweeping the sample past the secondary slit (essentially varying $x$ in Figure \ref{fig:3}), it is possible to obtain a profile of measurements analogous to those which can be measured using pixelated detectors in a transmission setup.\footnote{For time-of-flight instruments, it may be possible to achieve something equivalent using a collimator and imaging detector.}

Many of the algorithms already developed for Bragg-edge transmission tomography may be adapted to this geometry. Two significant differences exist;

\begin{enumerate}
    \item In this regime, the measurement direction is not aligned with the direction of a ray.
    \item The incident and diffracted beams are not collinear, meaning attenuation along the neutron ray must be considered.
\end{enumerate}

Attenuation provides a significant challenge --- each measurement now represents a weighted average of the strains along a ray. The contribution from each point along the ray is dependent on the total path length seen by a neutron diffracted at that point.

\subsection{Measurement Model}

Considering the effect of attenuation, we propose that diffraction tomography measurements can be modelled by a ray transform of the form:
\begin{equation}
\epsave = \frac{1}{\int_0^L e^{-\mu L_T(s)}\; \text{d}s} \int_0^L e^{-\mu L_T(s)} \kap^\Transp\bs{\epsilon}(s)\\\kap \; \text{d}s.
\label{eqn:aLRT}
\end{equation}

For a given measurement, strain in the $\kap$ direction is averaged along the ray with length $L$ as shown in Figure \ref{fig:3}. $L_T(s)$ is the total path length seen by a neutron diffracted at a point $s$ units along this ray (including the incident path). $\mu$ is the effective neutron attenuation coefficient of the material being measured.

\subsection{Algorithm Selection}

A number of algorithms \citep{,abbey09,abbey12,kirkwood15,wensrich16a,wensrich16b,hendriks2017,gregg2017axi,gregg2018resid,hendriks2018traction,jidling2018probabilistic} have recently been presented for transmission strain tomography, and most could be modified to suit the problem at hand.

Of the those presented thus far, the most promising algorithms have been those which model the unknown strain field as a Gaussian Process (GP). Compared to other approaches, the convergence of GP regression algorithms for strain reconstruction has generally been significantly more rapid, partially owing to their implicit implementation of the equilibrium constraint and explicit implementation of known boundary (loading) conditions \citep{hendriks2018traction}.

GP regression also presents a number of other benefits compared to other algorithms. Being a probabilistic approach, GPs are able to utilise the confidence of each measurement in the resulting reconstruction (i.e. by giving less weighting to more uncertain measurements). The GP technique also allows the confidence interval of the reconstruction to be calculated at every point.

Finally, the GP technique is non-parametric. This means that complexity does not increase with the resolution of the reconstruction. While a small, fixed number of so-called \emph{hyperparameters} influence the reconstruction, these can be optimised from the measurements alone using a likelihood maximisation method.

For these reasons, we will implement a GP reconstruction algorithm in this paper.

\subsection{Brief Introduction to Gaussian Processes}

GP regression is explained in detail in \citep{rasmussen2006gaussian}. Specifics related to implementing this technique for strain reconstruction can be found in \citep{hendriks2018traction,jidling2018probabilistic}.

Briefly, a GP is a machine learning technique that models an unknown field as a Gaussian random distribution of functions $\bs{f(x)}, \; \bs{x} \in \mathbb{R}^{\text{dim}(\bs{x})}$. 
This distribution is characterised by a mean function $\bs{m(x)}$ and covariance function $\bs{K(x,x')}$, where;
\begin{align*}
\bs{m(x)} &= \mathbb{E}\left[\bs{f(x)}\right] \\
\bs{K(x,x')} &= \mathbb{E}\left[(\bs{f(x)}-\bs{m(x)})(\bs{f(x')}-\bs{m(x')})^\Transp\right]
\end{align*}

The choice of covariance function $\bs{K(x,x')}$ can have a significant impact on the resulting reconstruction. For this application, we build a GP around the squared-exponential covariance function, which assumes a high degree of smoothness and has shown promise in modelling strain \citep{hendriks2019robust,hendriks2018traction,jidling2018probabilistic} --- a usually smooth phenomena;
\begin{equation*}
\bs{K(x,x')} = \sigma_f^2 \; \text{exp}\left(\frac{-\norm{\bs{x}-\bs{x'}}^2}{2\ell^2}\right).
\end{equation*}

Here, $\sigma_f^2$ is the variance on the prior (sometimes called the signal variance), and $\ell$ is a length-scale. These hyperparameters inform the most likely functions to be drawn from the GP.
For instance, a small length-scale favours rapidly changing reconstructions, while larger length-scales motivate smoother, slower-varying realisations.

Gaussian process regression involves estimating a function value at a user specified sample point $\bs{x_*}$ given a set of data, $\mathcal{D} = \left\{y_i,\bs{\eta}_i \; \vert \; \forall \; i=1,\dots,n\right\}$, where each measurement is of the form;
\begin{equation*}
y_i = \tL_{\bs{\eta}_i} \bs{f(x)} + e_i.
\end{equation*}

Here, $\tL_{\bs{\eta}_i} \bs{f(x)}$ is a linear transformation of the underlying function $\bs{f(x)}$ that is parametrised by the set $\bs{\eta}_i$. $e_i\sim \mathcal{N}(0,\sigma_i^2)$ is assumed zero-mean Gaussian noise with standard deviation $\sigma_i$.

Given that GPs are closed under linear operators \citep{papoulis2002probability,wahlstrom2015modeling}, the measurements $\bs{Y} = [y_1, y_2, \ldots y_n]^\Transp$ and a function value estimate $\hat{f}(\bs{x_*})$ are jointly Gaussian \citep{rasmussen2006gaussian};
\begin{equation*}
\begin{bmatrix}
\bs{Y} \\
\bs{\hat{f}(x_*)}
\end{bmatrix} \sim \mathcal{N} \left(
\begin{bmatrix}
\bs{\mu_y} \\
\bs{m(x_*)}
\end{bmatrix},
\begin{bmatrix}
\bs{K_{yy'}+\Sigma_m} & \bs{K_{y\hat{f}'}}\\
\bs{K_{\hat{f}y'}} & \bs{K(x_*,x_*)}
\end{bmatrix} 
\right)
\end{equation*}

Where $\bs{\Sigma_m}$ is a diagonal matrix in which the $i^{\text{th}}$ entry corresponds to the variance of the $i^{\text{th}}$ measurement, and the cross-covariance matrix $\bs{K_{y\hat{f}'}}$ and covariance matrix $\bs{K_{yy'}}$ are given by\footnote{note: $\bs{K_{\hat{f}y'}} = \bs{K_{y\hat{f}'}}^\Transp$}:

\begin{equation*}
    \bs{K_{y\hat{f}'}} = \begin{bmatrix}
        \tL_{\bs{\eta}_1} \bs{K}(\bs{x},\bs{x_*}) \\
        \vdots \\
        \tL_{\bs{\eta}_n} \bs{K}(\bs{x},\bs{x_*})
    \end{bmatrix} 
\end{equation*}

and

\begin{equation*}
\begingroup 
\setlength\arraycolsep{1pt}
    \bs{K}_{\bs{y}\bs{y}'} = \begin{bmatrix}
        \tL_{\bs{\eta}_1} \bs{K}(\bs{x},\bs{x_*})\tL_{\bs{\eta}_1} '^\Transp & \cdots & \tL_{\bs{\eta}_1} \bs{K}(\bs{x},\bs{x_*})\tL_{\bs{\eta}_n}'^\Transp  \\
         \vdots & \ddots & \vdots \\ 
         \tL_{\bs{\eta}_n} \bs{K}(\bs{x},\bs{x_*})\tL_{\bs{\eta}_1} '^\Transp & \cdots & \tL_{\bs{\eta}_n} \bs{K}(\bs{x},\bs{x_*})\tL_{\bs{\eta}_n} '^\Transp \\ 
    \end{bmatrix}.
\endgroup
\end{equation*}

Here, the notation $\tL'$ is used to distinguish a linear transform operating on $\bs{f(x')}$ from a linear transform $\tL$ operating on $\bs{f(x)}$.
Finally, we can condition the prior estimate $f(\bs{x_*})$ on the known measurement values to give a posterior estimate with mean and variance obtained from;
\begin{align*}
        \bs{\mu}_{\bs{f}_*|\bs{Y}} &= \bs{m}(\bs{x}_*) + \bs{K_{\hat{f}y'}}\left(\bs{K}_{\bs{yy}'}+\bs{\Sigma_m} \right)^{-1}(\bs{Y}-\bs\mu_y), \\
        \Sigma_{\bs{f}_*|\bs{Y}}  &= \bs{K}(\bs{x}_*,\bs{x}_*) - \bs{K_{\hat{f}y'}}\left(\bs{K}_{\bs{yy}'}+\bs{\Sigma_m}\right)^{-1}\bs{K_{y\hat{f}'}}.
\end{align*}

\subsection{A Gaussian Process for Diffraction Tomography}

In this paper, we develop an algorithm for axisymmetric, 2D systems. This choice is motivated by the simplifications afforded by geometry and the reduced measurement time required for validation in this proof-of-concept study.

Conceptually, there is well defined path to extend this method to arbitrary 2D and even 3D systems. Practical issues may arise --- these are discussed in Section \ref{sec:ext}.

In the process, we also contribute a GP for strain which leverages the strong constraint provided by axisymmetry --- this formulation may be useful outside this specific problem.

We assume a 2D circular sample of radius $R$ with origin at it's centre and are concerned with reconstructing the tensor strain distribution, $\bs{\epsilon}$, within this sample. Naturally, we will work in polar coordinates $(r,\theta)$. Under an axisymmetry assumption, the strain tensor can be written in terms of two nonzero in-plane components which vary only in the radial direction: $\bs{\epsilon}(r) = \begin{bmatrix} \epsilon_{rr}(r) & \epsilon_{\theta \theta}(r) \end{bmatrix}^\Transp.$

Our problem is constrained by equilibrium, which is described in terms of the stress tensor, $\bs{\sigma}(r) = \begin{bmatrix} \sigma_{rr}(r) & \sigma_{\theta \theta}(r) \end{bmatrix}^\Transp$ --- a linear transformation of strain --- in polar coordinates by the differential equation:

\begin{equation}
\pd{\sigma_{rr}}{r} +  \frac{1}{r}\bigg(\sigma_{rr}(r)-\sigma_{\theta \theta}(r)\bigg)=0.
\label{eq:eql}
\end{equation}

To encode this constraint, we define stress via a scalar potential $\phi(r)$ through the following mapping:

\begin{equation*}
\bs{\sigma}(r) = \begin{bmatrix} \sigma_{rr}(r) \\ 
\sigma_{\theta \theta}(r) \end{bmatrix} = \begin{bmatrix} \phi(r) \\ 
r \pd{\phi}{r} + \phi(r) \end{bmatrix}
\end{equation*}

By relating the components of stress in this way, any stress field resulting from the potential $\phi(r)$ automatically satisfies equilibrium\footnote{Direct substitution into Equation \ref{eq:eql} confirms this.}.

This approach closely resembles the Airy stress function technique first encoded in a GP in \citep{jidling2018probabilistic}, but avoids the singularity near $r=0$ present in the Airy stress mapping for polar coordinates. The mapping chosen here also encodes a key constraint that arises from the axisymmetric polar coordinate system; $\sigma_{rr}(0) = \sigma_{\theta \theta}(0)$.

In this paper, we develop our algorithm assuming plane-stress\footnote{A minor modification is required for plane-strain.}, for which Hooke's Law takes the form:
\begin{align*}
\bs{\epsilon}(r) = \begin{bmatrix} \epsilon_{rr}(r) \\ 
\epsilon_{\theta \theta}(r) \end{bmatrix} = \frac{1}{E}\begin{bmatrix} \sigma_{rr}(r)- \nu \sigma_{\theta \theta}(r) \\ 
-\nu\sigma_{rr}(r) + \sigma_{\theta \theta}(r) \end{bmatrix}
\end{align*}

We then have the following relationships between the scalar potential for which we build our GP, $\phi(r)$, and the strains we wish to estimate:

\begin{equation*}
\bs{\epsilon}(r) = \frac{1}{E}\begin{bmatrix} (1-\nu) -\nu r \pd{}{r} \\ (1-\nu) + r \pd{}{r} \end{bmatrix}\phi(r). \end{equation*}

To improve numerical stability, we neglect the scaling factor $1/E$ and relate the potential to the strains through the mapping $\bs{H}$ by:

\begin{equation*}
\bs{\epsilon}(r) = \begin{bmatrix} (1-\nu) -\nu r \pd{}{r} \\ (1-\nu) + r \pd{}{r} \end{bmatrix}\phi(r) =  \bs{H} \phi(r). \end{equation*}

The strain field is mapped to a measurement through the measurement model:
\begin{equation*}
\epsave = \begin{bmatrix} W_\mu \int_0^{L} w_\mu(s)w_r(s) (\cdot) \; \text{d}s  \\ W_\mu \int_0^{L} w_\mu(s)w_\theta(s) (
\cdot) \; \text{d}s \end{bmatrix}^\Transp \bs{\epsilon}(r) = \bs{M} \bs{\epsilon}(r).
\end{equation*}

With weights $W_\mu, w_\mu(s), w_r(s)$ and $w_\theta(s)$ as defined in Appendix \ref{app:A}. These mappings $\bs{H}$ and $\bs{M}$ together comprise the linear transformation $\tL_\eta $ that acts on $\phi(r)$ to provide a measurement $\epsave$:
\begin{align*}
\epsave = \bs{M} \bs{\epsilon} = \bs{MH} \phi(r) = \tL_\eta \phi(r)
\end{align*}

\subsection{Hyperparameter Selection}

The GP hyperparameters $\bs{h}=\begin{bmatrix}\ell & \sigma_f \end{bmatrix}^\Transp$ can have a significant impact on the fidelity of a reconstruction. Selection of these hyperparameters does not require \emph{a-priori} knowledge of the system, or user intervention --- they can be determined from the measurements alone.

This is achieved by posing an optimisation problem to maximise the marginal likelihood of the measurements given a set of hyperparameters $\bs{h}$ \cite{rasmussen2006gaussian}. In practice, for numerical stability, the hyperparameters are estimated by maximising the log marginal likelihood:
\begin{equation*}
\argmax{\bs{h}} \Big(\log\text{det}(\bs{K}_{\bs{yy}}(\bs{h})+ \bs{\Sigma}_m) -\bs{y}^\Transp(\bs{K}_{\bs{yy}} (\bs{h})+ \bs{\Sigma}_m)^{-1}\bs{y} \Big).
\end{equation*}

There may be several local minima for a given set of measurements. These can correspond to different interpretations of the data. For instance, the same data may involve noisy measurements of a smooth function or precise measurements of a quickly-varying function. Use of a multi-start optimisation procedure and/or non-gradient based method (e.g. simulated annealing) can help to avoid these local minima and find the hyperparameters that are most likely given a set of measurements.

\section{Demonstration}
\subsection{Simulation}
We first demonstrate our algorithm for the axisymmetric strain field previously examined in \citep{gregg2017axi}:
\begin{equation*}
\begin{bmatrix} \epsilon_{rr}(r) \\ \epsilon_{\theta\theta}(r)
\end{bmatrix}  = e_0
\begin{bmatrix} \frac{(7+5\nu)R^2 + (1+\nu)(9r^2-16Rr)}{12R^2} -\left(1-\frac{r}{R} \right)^2 \\
\frac{(7+5\nu)R^2 + (1+\nu)\left(3r^2-8Rr\right)}{12R^2} -\left(1-\frac{r}{R} \right)^2
\end{bmatrix},
\end{equation*}
where $e_0$ is a scaling factor and $R$ the sample radius. 

This field satisfies equilibrium but not compatibility (i.e. is residual), and corresponds to the elastic component of strain set up in response to a hydrostatic eigenstrain of the form $\epsilon^*_{rr}(r)=\epsilon^*_{\theta \theta}(r) = e_0(1-\tfrac{r}{R})^2.$ The method by which this field was generated is described in detain in the appendix of \citep{gregg2017axi}.

A total of 15 diffraction tomography measurements were simulated across the width of a sample with radius $R=6.5$mm via Equation \ref{eqn:aLRT} to form the profile shown in Figure \ref{fig:5}(a). Material properties and attenutation characteristics of steel were chosen: $E = 220$ GPa, $\nu = 0.3$, $\mu=120$ m$^{-1}$. Measurements were corrupted by simulated mean-zero Gaussian measurement noise with standard deviation $\sigma=0.5\times10^{-4}$ --- typical of that which can be expected on a diffraction strain scanner given gauge volumes of this size and sensible measurement times \citep{kirstein2009strain}.

Hyperparameters $\bs{h} = \begin{bmatrix}\ell & \sigma_f \end{bmatrix}^\Transp$ of the covariance function were determined by the likelihood maximisation procedure described earlier and were found to be 11.2mm and $6.7\times10^{-3}$, respectively.

Reconstruction results are shown in Figure \ref{fig:5}(b) and (c), and show close agreement with the true field, which lies within the 1-$\sigma$ confidence interval of the reconstruction over it's entire span.

The reconstruction differs most significantly from the true field near the sample centre (at small $r$). The confidence interval of the result is also widest here. The cause seems to be twofold:

\begin{enumerate}
    \item Strains near $r=0$ do not contribute at all to the majority of measurements as only rays passing close to the sample centre can see these strains (the smallest $r$ seen by a given ray is $|x|$).
    \item Those rays which do manage to probe strains at small $r$ do so with large path lengths and consequently the relative weighting of points near $r=0$ in the average is small. 
\end{enumerate}

 The reconstruction was seen to converge to the true field as additional measurements were added near $x=0$ and as less attenuation was simulated.

 Overall, this simulation shows a similar quality reconstruction as \cite{gregg2017axi}, however it should be noted that this algorithm required far fewer measurements (15 vs 512) to achieve this result. This is largely owed to the intrinsic equilibrium constraint encoded in the GP.

\begin{figure}
\begin{center}
\begin{overpic}[width=0.8\linewidth]{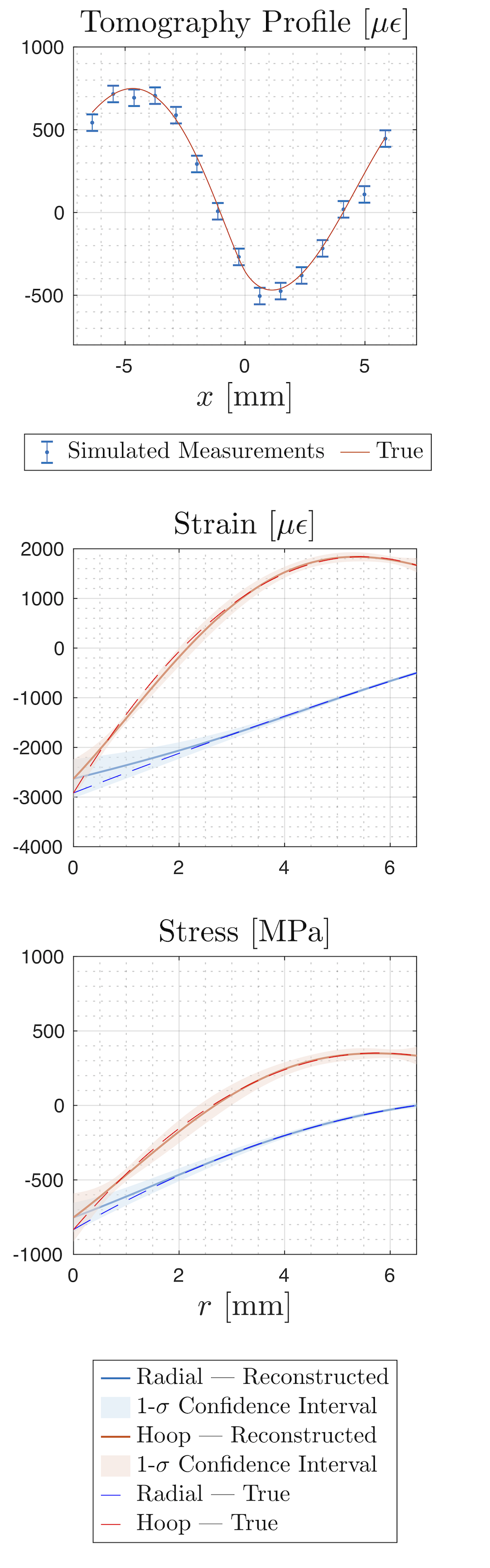}
 \put (28,87) {\Large$\displaystyle\text{(a)}$}
 \put (28,55) {\Large$\displaystyle\text{(b)}$}
 \put (28,29) {\Large$\displaystyle\text{(c)}$}
\end{overpic} \\ \ \\
\vspace{-0.8cm}
\caption{(a) Simulated Tomography profile. (b) Reconstructed and true strain fields. (c) Corresponding stress fields.}
\vspace{-0.8cm}
\label{fig:4}
\end{center}
\end{figure}

\FloatBarrier

\subsection{Experiment}

Following success in simulation, the reconstruction algorithm was applied to data collected on the KOWARI constant wavelength diffractometer within the Australian Centre for Neutron Scattering (ACNS) at the Australian Nuclear Science and Technology Organisation (ANSTO) \citep{kirstein2009strain,kirstein2010kowari}.

\subsubsection*{Sample Design}

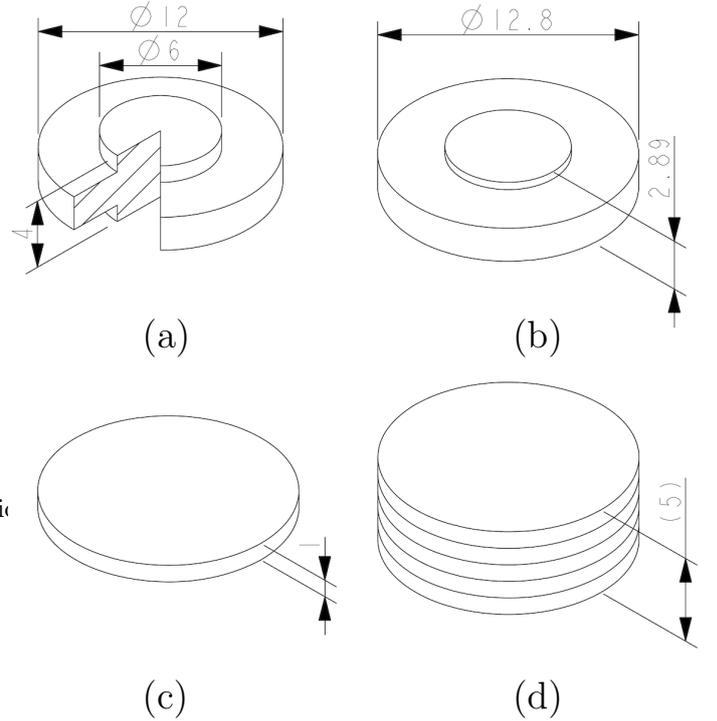
\begin{figure}
\begin{center}
\begin{overpic}[width=0.5\textwidth]{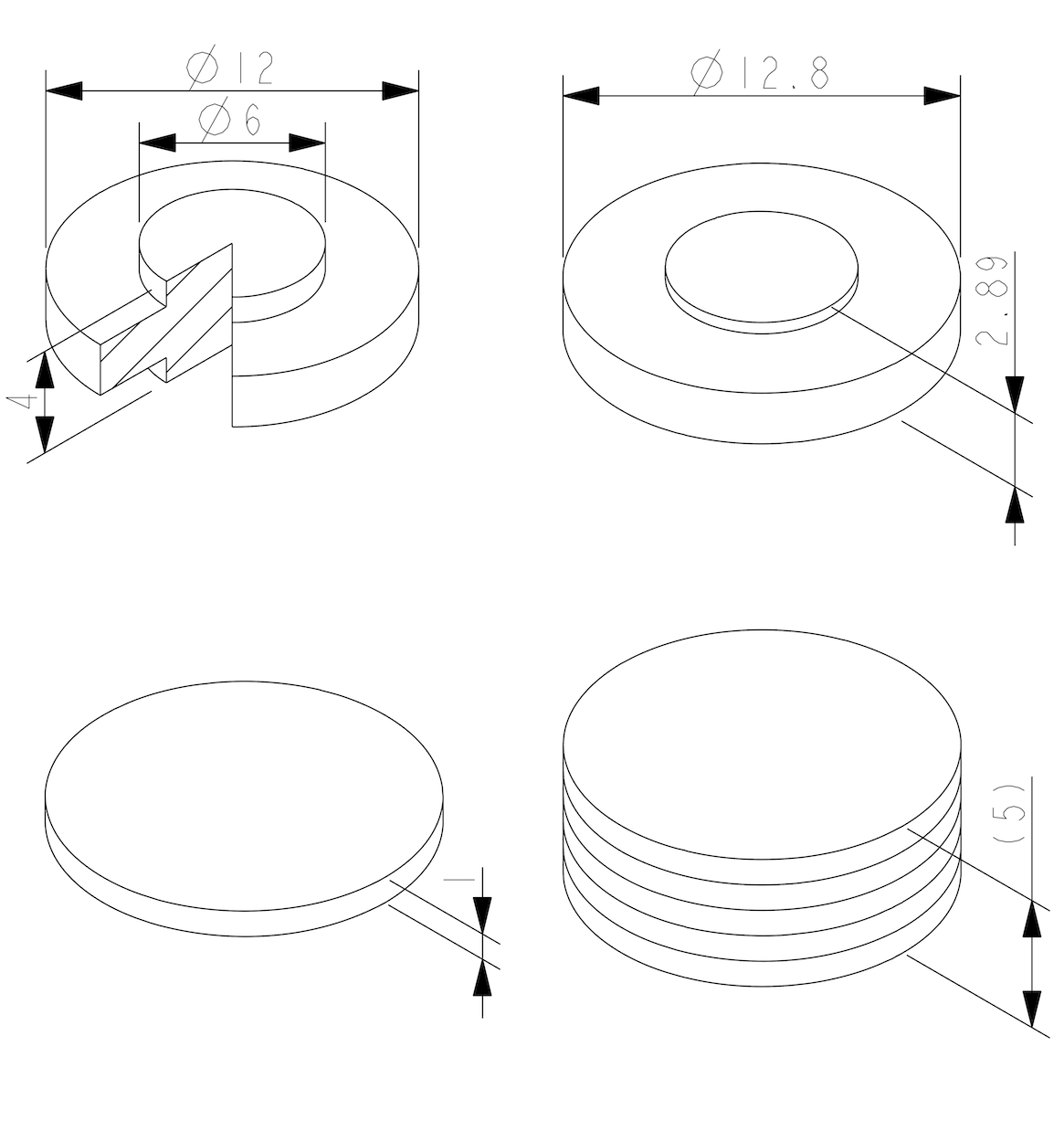}
 \put (18,49) {\Large$\displaystyle\text{(a)}$}
 \put (68,49) {\Large$\displaystyle\text{(b)}$}
 \put (18,0) {\Large$\displaystyle\text{(c)}$}
 \put (68,0) {\Large$\displaystyle\text{(d)}$}
\end{overpic} \\ \ \\
\vspace{-0.3cm}
\caption{Sample dimensions after (a) Initial machining, (b) Deformation, (c) Final machining, and (d) Final assembly.}
\vspace{-0.8cm}
\label{fig:5}
\end{center}
\end{figure}

The sample consisted of a stack of small aluminium discs, which, with respect to Figure \ref{fig:5}, were manufactured as follows.

\begin{enumerate}[label=(\alph*)]
    \item Five stepped discs were manufactured from $D\approx 12$mm diameter 7075-T6 aluminium round bar with dimensions as shown.
    \vspace{-0.2cm}
    \item These discs were then individually crushed in a deformation-controlled process to a uniform thickness of $2.89\pm0.02$ mm.
        \vspace{-0.2cm}
    \item The crushed discs were then each symmetrically faced to a thickness of $t=1$ mm.
        \vspace{-0.2cm}
    \item Finally, the discs were stacked and glued together to form one sample.
    \vspace{-0.2cm}
\end{enumerate}

The final dimensions of each disc are in line with the typical rule-of-thumb for plane stress systems ($D/t \approx 10$) \citep{beer2010mechanics}. Stacking was essential to provide sufficient material for both validation and diffraction tomography measurements to be performed in the limited available beamtime.
\subsubsection*{Calibration and Experimental Setup}

Initial work centred on characterising the beam and finding an optimal focussing condition for the monochromator on KOWARI.

Intensity variation across the beam introduces an additional non-uniform weighting to the measured average strain along a ray. The intensity profile was characterised using an iron-powder standard sample and tracking the Fe (110) reflection. The position of this peak is similar to that of the Al (200) reflection used for the tomography measurements, however Fe provided faster measurements.

By adjusting the monochromator focus, variation in this profile was minimised. The final condition achieved less than 10\% variation across the sample and was assumed constant.

Nominally 90-degree geometry was adopted (89.6$^\circ$), and after line scans to find the sample edges, 11 diffraction tomography measurements were taken across the width of the sample using a $w=1$ mm secondary slit width ($1\times L \times 5$ mm gauge volumes). Measurements were performed until 5000 counts had been recorded at each point. Total measurement time was approximately 12 hours.

Validation measurements of the hoop, radial and axial components of strain were performed using a $1\times1\times 5$ mm gauge volume and tracking the Al (311) reflection and were also performed over 12 hours. Results from the two experiments were correlated using a standard powder sample which was measured under both regimes. The unstrained lattice parameter $d_0$ was calculated from the three measured strain components using the plane-stress assumption.

\subsubsection*{Results}

\begin{figure}
\begin{center}
\begin{overpic}[width=0.8\linewidth]{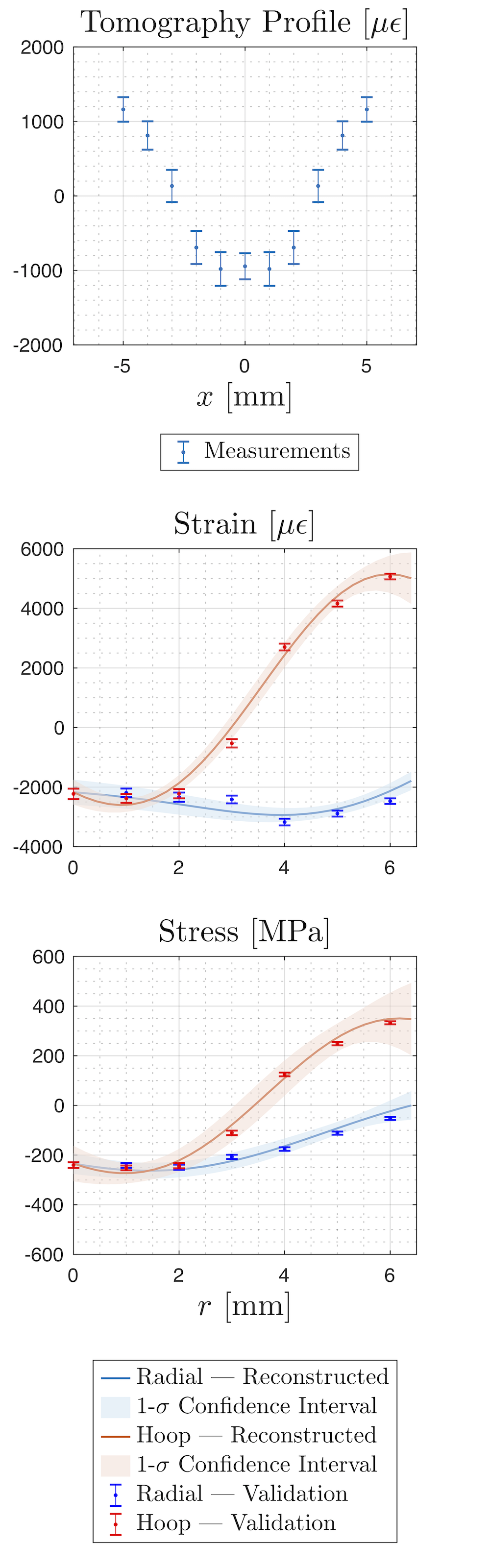}
 \put (28,87) {\Large$\displaystyle\text{(a)}$}
 \put (28,55) {\Large$\displaystyle\text{(b)}$}
 \put (28,29) {\Large$\displaystyle\text{(c)}$}
\end{overpic} \\ \ \\
\vspace{-0.8cm}
\caption{(a) Measured Tomography profile, (b) Reconstructed strain fields against validation measurements, (c) Corresponding stresses.}
\vspace{-0.8cm}
\label{fig:6}
\end{center}
\end{figure}

Reconstruction results are shown in Figure \ref{fig:6}. In general, the reconstruction shows excellent agreement with the validation measurements --- with almost all measurements lying within the 1-$\sigma$ confidence interval of the reconstruction.

Quantitatively, the difference between the tomographic reconstruction and validation measurements was Gaussian, with mean $\approx-5.5\times10^{-5}$ and standard deviation $\approx2.8\times10^{-4}$ --- respectively two and one order of magnitude less than the strains under examination. This suggests that the difference is largely due to unrejected measurement noise and not a systematic bias. Note that the systematic error near $r=0$ due to attenuation seen in the simulated example are not present here --- aluminium has an attenuation coefficient nearly ten times smaller than steel.

\FloatBarrier

\section{Discussion and Potential Future Extensions}
\label{sec:ext}

This proof-of-concept demonstration leaves many potential improvements to the technique on the table for future implementation:

\begin{enumerate}
    \item A sensitivity analysis was performed as to the effect of secondary-slit width on simulated tomography profiles. It was found that the variation was minimal up to slit widths of 2 mm for some smooth strain fields. However, this effect may be more important for discontinuous fields (e.g. ring-and-plug) or regions of high strain gradient. Future implementations of this technique could implement slit width into the measurement model, treating each measurement as an area rather than line integral.
    \item Intensity variation across the beam could be normalised in this case, but might be unavoidable at other neutron sources. Further improvements to the reconstruction algorithm could be made by implementing a weighting due to intensity in the measurement model. This weighting could either then be measured at the source in question or determined via hyperparameter optimisation.
    \item Intensity variation due to sample texture was also not found to play a significant role in this experiment, but again could be implemented as an additional weighting in the measurement model and either measured or optimised for.
    \item For the validation measurements, $d_0$ was calculated from the three measured strain components and the plane-stress condition. An average, constant $d_0$ was assumed for the tomography measurements. In some cases --- particularly with unannealed/preprocessed samples, $d_0$ variation can result in significant pseudostrains. It is possible to estimate $d_0$ alongside the strains using the GP technique and such an extension to the reconstruction algorithm may be useful in some cases --- see \citep{hendriks2019robust} for more information.
\end{enumerate}

Extension of the algorithm presented here to 2D and 3D strain fields is conceptually straightforward, though practical issues may arise. A 2D implementation could involve taking projections of a sample by sweeping it past the secondary slit, then rotating the sample about it's centre and repeating the process. A 2D reconstruction algorithm would likely resemble that in \citep{hendriks2018traction}, with modifications made for attenuation and measurement direction.

While the results of this study found that diffraction tomography reconstructions can achieve results in agreement with conventional `pointwise' measurements in equal or less beamtime, further work is required to show that this technique is viable for 2D systems, particularly with larger samples where attenuation of the beam may prove a limiting factor.

That being said, preliminary results suggest that --- particularly for better scattering materials such as steel --- projections of a similar confidence could be had in as little as 2 hours, compared to 7 hours to collect three components with conventional pointwise scanning (using a 1 mm slit width in both cases). 

If a method for collecting entire projections at once is developed (e.g. using parallel collimators and a pixelated detector on a time-of-flight instrument), projections could be obtained in the time it presently takes to perform a single diffraction tomography measurement.

One of the more promising outcomes of this demonstration is the prospect of combined neutron transmission-diffraction tomography. In this case, both measurements could be performed concurrently and then processed by single reconstruction algorithm that takes into account their relative uncertainty.

Extension of this method to 2D systems, time-of-flight instruments and integration with existing transmission reconstruction algorithms forms a natural basis for future work.

\section{Conclusion}

In this paper, we proposed a new technique for strain measurement that we call neutron diffraction tomography. This technique draws on the recent activity in the field of Bragg-edge transmission strain tomography and aims to achieve similar results using conventional strain scanning geometry and instrumentation.

Reconstructions from simulated and experimental data were achieved and showed close agreement with validation measurements for axisymmetric systems with similar beamtime requirements.

We believe the method is viable. Future work involves extension to arbitrary 2D systems, where it may also be valuable in a combined transmission-diffraction strain reconstruction algorithm.

\section{Acknowledgements}
This work was supported by the Australian Research Council through a Discovery Project Grant No. DP17010 2324. Access to the KOWARI instrument was made possible through the ANSTO user-access program (Proposal No. 7318). The authors would also like to thank AINSE Limited for providing financial assistance (PGRA) and support to enable work on this project.

\appendix
\section{Measurement Model Derivations}
\label{app:A}

With respect to the coordinate system shown in Figure \ref{fig:3}, we propose a measurement model of the form:
\begin{equation*}
\epsave = \frac{1}{\int_0^L e^{-\mu L_T(s)}\; \text{d}s} \int_0^L e^{-\mu L_T(s)} \kap^\Transp\bs{\epsilon}(s)\kap \; \text{d}s, \\
\end{equation*}
where, $\kap^\Transp\bs{\epsilon}(s)\kap$ provides the average normal component of strain aligned with the measurement direction $\kap$ in Cartesian coordinates by:
\begin{equation*}
\kap^\Transp\bs{\epsilon}(s)\kap = \kappa_x^2 \epsilon_{xx} + 2\kappa_x\kappa_y \epsilon_{xy} + \kappa_y^2 \epsilon_{yy}.
\end{equation*}

We assume that the $y$ axis of our right-handed coordinate system is aligned with the path of diffracted neutron rays (in the direction of the detector). Thus, for a given diffraction angle $2\vartheta$, we have a measurement direction vector $\kap$ given by:
\begin{equation*}
\kap = \begin{bmatrix}\cos (\pi/2 - \vartheta) & \sin (\pi/2 - \vartheta) \\ -\sin (\pi/2 - \vartheta) & \cos (\pi/2 - \vartheta) \end{bmatrix}
\begin{bmatrix} 0 \\ 1 \end{bmatrix} = \begin{bmatrix}\cos \vartheta \\ \sin \vartheta \end{bmatrix}.
\end{equation*} \\
Converting strains from Cartesian to polar coordinates gives:
\begin{align*}
\kap^\Transp\bs{\epsilon}(s)\kap &= \cos^2\vartheta\bigg( \epsilon_{rr}(r) \cos^2{\theta} + \epsilon_{\theta \theta}(r) \sin^2{\theta} \bigg) \\
&\qquad+ 2\sin\vartheta\cos\vartheta \bigg(\sin{\theta}\cos{\theta}(\epsilon_{rr}(r)-\epsilon_{\theta \theta}(r)) \bigg) \\
&\qquad\qquad+ \sin^2\vartheta \bigg( \epsilon_{rr}(r) \sin^2{\theta} + \epsilon_{\theta \theta}(r) \cos^2{\theta} \bigg) \\
&= \epsilon_{rr}(r) \cos^2(\vartheta - \theta) + \epsilon_{\theta \theta}(r) \sin^2(\vartheta - \theta)
\end{align*}

In terms of the two nonzero in-plane components of strain in polar coordinates, we can then write:
\begin{equation*}
\epsave \approx W_\mu \int_0^{L} w_\mu(s) \bigg(w_r(s)\epsilon_{rr}(r) + w_\theta(s) \epsilon_{\theta \theta}(r)\bigg) \; \text{d}s
\end{equation*}

Where:
\begin{align*}
r=r(s) = \sqrt{x^2 + (s-\sqrt{2pR-p^2})^2}
\end{align*}

and:
\begin{align*}w_\mu(s) &= e^{-\mu L_T(s)} \\
W_\mu &= \frac{1}{\int_0^L w_\mu(s) \; \text{d}s} = \frac{1}{\int_0^L e^{-\mu L_T(s)} \; \text{d}s}\\
w_r(s) &= \cos^2(\vartheta-\theta)\\
w_\theta(s) &= \sin^2(\vartheta-\theta)\\
\end{align*}

The measurement can be separated into two integrals:
\begin{align*}
\epsave =& W_\mu \bigg( \int_0^{L} w_\mu(s)w_r(s)\epsilon_{rr}(r) \; \text{d}s + \int_0^{L} w_\mu(s)w_\theta(s) \epsilon_{\theta \theta}(r) \; \text{d}s \bigg)
\end{align*}

And written as a linear transformation of the strain tensor:
\begin{align*}\epsave =& \begin{bmatrix} W_\mu \int_0^{L} w_\mu(s)w_r(s) (\cdot) \; \text{d}s &W_\mu \int_0^{L} w_\mu(s)w_\theta(s) (
\cdot) \; \text{d}s \end{bmatrix} \begin{bmatrix} \epsilon_{rr} \\ \epsilon_{\theta \theta} \end{bmatrix}
\end{align*}

For future working, we define this transform $\bs{M}$:

\begin{equation*}
\epsave = \bs{M} \bs{\epsilon}.
\end{equation*}

\bibliography{references}

\begin{thebibliography}{10}
\expandafter\ifx\csname url\endcsname\relax
  \def\url#1{\texttt{#1}}\fi
\expandafter\ifx\csname urlprefix\endcsname\relax\def\urlprefix{URL }\fi
\expandafter\ifx\csname href\endcsname\relax
  \def\href#1#2{#2} \def\path#1{#1}\fi

\bibitem{noyan87}
I.~C. Noyan, J.~B. Cohen, Determination of strain and stress fields by
  diffraction methods, in: Residual Stress, Springer, 1987, pp. 117--163.

\bibitem{wensrich14}
C.~M. Wensrich, E.~H. Kisi, V.~Luzin, U.~Garbe, O.~Kirstein, A.~L. Smith, J.~F.
  Zhang, Force chains in monodisperse spherical particle assemblies:
  Three-dimensional measurements using neutrons, Phys. Rev. E 90 (2014) 042203.
\newblock \href {http://dx.doi.org/10.1103/PhysRevE.90.042203}
  {\path{doi:10.1103/PhysRevE.90.042203}}.

\bibitem{abbey09}
B.~Abbey, S.~Y. Zhang, W.~J. Vorster, A.~M. Korsunsky, Feasibility study of
  neutron strain tomography, Procedia Engineering 1~(1) (2009) 185--188.

\bibitem{abbey12}
B.~Abbey, S.~Y. Zhang, W.~Vorster, A.~M. Korsunsky, Reconstruction of
  axisymmetric strain distributions via neutron strain tomography, Nuclear
  Instruments and Methods in Physics Research Section B: Beam Interactions with
  Materials and Atoms 270 (2012) 28--35.

\bibitem{kirkwood15}
H.~J. Kirkwood, S.~Y. Zhang, A.~S. Tremsin, A.~M. Korsunsky, N.~Baimpas,
  B.~Abbey, Neutron strain tomography using the radon transform, Materials
  Today: Proceedings 2 (2015) S414--S423.

\bibitem{wensrich16a}
C.~M. Wensrich, J.~Hendriks, M.~H. Meylan, {B}ragg edge neutron transmission
  strain tomography in granular systems, Strain 52~(1) (2016) 80--87.

\bibitem{wensrich16b}
C.~Wensrich, J.~Hendriks, A.~Gregg, M.~Meylan, V.~Luzin, A.~Tremsin,
  {B}ragg-edge neutron transmission strain tomography for in situ loadings,
  Nuclear Instruments and Methods in Physics Research B: Beam Interactions with
  Materials and Atoms 383 (2016) 52--58.

\bibitem{hendriks2017}
J.~N. Hendriks, A.~W.~T. Gregg, C.~M. Wensrich, A.~S. Tremsin, T.~Shinohara,
  M.~Meylan, E.~H. Kisi, V.~Luzin, O.~Kirsten, Bragg-edge elastic strain
  tomography for in situ systems from energy-resolved neutron transmission
  imaging, Physical Review Materials 1~(5) (2017) 053802.

\bibitem{gregg2017axi}
A.~W.~T. Gregg, J.~N. Hendriks, C.~M. Wensrich, M.~H. Meylan, Tomographic
  reconstruction of residual strain in axisymmetric systems from bragg-edge
  neutron imaging, Mechanics Research Communications 85 (2017) 96 -- 103.

\bibitem{gregg2018resid}
A.~W.~T. Gregg, J.~N. Hendriks, C.~M. Wensrich, A.~Wills, A.~S. Tremsin,
  V.~Luzin, T.~Shinohara, O.~Kirsten, M.~H. Meylan, E.~H. Kisi, Tomographic
  reconstruction of two-dimensional residual strain fields from bragg-edge
  neutron imaging, Physical Review Applied.

\bibitem{hendriks2018traction}
J.~N. Hendriks, A.~W.~T. Gregg, C.~M. Wensrich, A.~Wills, Implementation of
  traction constraints in bragg-edge neutron transmission strain tomography,
  arXiv preprint arXiv:1805.09760.

\bibitem{jidling2018probabilistic}
C.~Jidling, J.~N. Hendriks, N.~Wahlstr{\"o}m, A.~W.~T. Gregg, T.~B. Sch{\"o}n,
  C.~M. Wensrich, A.~Wills, Probabilistic modelling and reconstruction of
  strain, Nuclear Instruments and Methods in Physics Research Section B: Beam
  Interactions with Materials and Atoms 436 (2018) 141 -- 155.
\newblock \href {http://dx.doi.org/https://doi.org/10.1016/j.nimb.2018.08.051}
  {\path{doi:https://doi.org/10.1016/j.nimb.2018.08.051}}.

\bibitem{santisteban02b}
J.~Santisteban, L.~Edwards, M.~Fitzpatrick, A.~Steuwer, P.~Withers, M.~Daymond,
  M.~Johnson, N.~Rhodes, E.~Schooneveld, Strain imaging by bragg edge neutron
  transmission, Nuclear Instruments and Methods in Physics Research Section A:
  Accelerators, Spectrometers, Detectors and Associated Equipment 481~(1)
  (2002) 765--768.

\bibitem{luzin19}
V.~Luzin, A.~W.~T. Gregg, J.~N. Hendriks, C.~M. Wensrich, Neutron diffraction
  strain tomography: 2d axisymmetric sample geometry, submitted to Strain.

\bibitem{rasmussen2006gaussian}
C.~E. Rasmussen, C.~K. Williams, Gaussian processes for machine learning,
  Vol.~1, MIT press Cambridge, 2006.

\bibitem{hendriks2019robust}
J.~N. Hendriks, C.~M. Wensrich, A.~Wills, V.~Luzin, A.~W.~T. Gregg, Robust
  inference of two-dimensional strain fields from diffraction-based
  measurements, Nuclear Instruments and Methods in Physics Research Section B:
  Beam Interactions with Materials and Atoms 444 (2019) 80--90.

\bibitem{papoulis2002probability}
A.~Papoulis, S.~U. Pillai, Probability, random variables, and stochastic
  processes, Tata McGraw-Hill Education, 2002.

\bibitem{wahlstrom2015modeling}
N.~Wahlstr{\"o}m, Modeling of magnetic fields and extended objects for
  localization applications, Ph.D. thesis, Link{\"o}ping University Electronic
  Press (2015).

\bibitem{kirstein2009strain}
O.~Kirstein, V.~Luzin, U.~Garbe, The strain-scanning diffractometer kowari,
  Neutron News 20~(4) (2009) 34--36.

\bibitem{kirstein2010kowari}
O.~Kirstein, U.~Garbe, V.~Luzin, Kowari-opal's new stress diffractometer for
  the engineering community: Capabilities and first results, in: Materials
  Science Forum, Vol. 652, Trans Tech Publ, 2010, pp. 86--91.

\bibitem{beer2010mechanics}
F.~Beer, E.~Johnston~Jr, J.~Dewolf, D.~Mazurek, Mechanics of materials, sixth
  edit edition (2010).

\end{thebibliography}
\end{document}